\documentclass[traditabstract]{aa} 
\usepackage{graphics}

\def\la{\lower.5ex\hbox{$\; \buildrel < \over \sim \;$}}
\def\ga{\lower.5ex\hbox{$\; \buildrel > \over \sim \;$}}

\begin{document}      
   
   \title{The SPECFIND V2.0 catalogue of radio cross-identifications and spectra.}
   \subtitle{SPECFIND meets the Virtual Observatory}

   \author{B.~Vollmer, B.~Gassmann, S.~Derri\`ere, T.~Boch, M.~Louys, F.~Bonnarel, P.~Dubois, F.~Genova, F.~Ochsenbein}

   \offprints{B.~Vollmer, e-mail: bvollmer@astro.u-strasbg.fr}

   \institute{CDS, Observatoire astronomique de Strasbourg, UMR 7550, 11, rue de l'universit\'e, 
     67000 Strasbourg, France}

   \date{Received / Accepted}

   \authorrunning{B.~Vollmer et al.}
   \titlerunning{The SPECFIND V2.0 catalogue}

\abstract{
  The new release of the SPECFIND radio cross-identification catalogue,
  SPECFIND V2.0, is presented. It contains 107488 cross-identified objects with at least 
  three radio sources observed at three independent frequencies.
  Compared to the previous release the number of entry radio catalogues is increased 
  from 20 to 97 containing 115 tables. This large increase was only made possible
  by the development of four tools at CDS which use the standards and infrastructure of the
  Virtual Observatory (VO). This was done in the framework of the VO-TECH European Design Study
  of the Sixth Framework Program.
  We give an overview of the different classes of radio sources that a user can encounter.
  Due to the increase of frequency coverage of the input radio catalogues,
  this release demonstrates that the SPECFIND algorithm is able to detect spectral 
  breaks around a frequency of $\sim 1$~GHz.
}

\maketitle

\section{Introduction \label{sec:introduction}}

The cross-identification of radio sources observed in the centimeter to meter wavelength domain
with different instruments is a rather difficult task, because of huge differences in sensitivity,
spatial resolution, and the non-simultaneous observations of variable sources. 
Especially the very different spatial resolutions of single dish telescopes
and interferometers are difficult to handle.
On the other hand, most sources show a power-law spectral energy
distribution in these wavelengths due to synchrotron or thermal emission. 
Synchrotron emission produces a power law spectrum with a possible cut-off
or reversal of the spectral index at low frequencies due to self-absorption or 
comptonization. The spectrum of thermal electrons is flat in the optically thin domain.

In Vollmer et al. (2005a) we presented the SPECFIND tool for the extraction
of cross-identifications and radio continuum 
spectra from radio source catalogues contained in the VizieR database of the Centre de Donn\'ees
astronomiques de Strasbourg (CDS).
The SPECFIND cross-identification tool takes advantage of the power-law shape of the
spectra. In addition, it takes into account the angular resolution of the observations, the source
size, and the flux densities observed at a given frequency. The SPECFIND tool also insures that a radio source 
cannot be assigned to more than one physical object.

In a first release (SPECFIND V1.0; Vollmer et al. 2005a,b)
we cross-identified the sources of the 20 largest radio catalogues in VizieR
(Ochsenbein et al. 2000), representing 3.5 million sources. Our work led to more than 
$700\,000$ independent cross-identifications between sources from different radio catalogues
and $\sim 67\,000$ independent radio spectra with more than two independent frequencies.  

The information contained in radio catalogues is heterogeneous containing different entries,
(e.g. peak flux or integrated flux) and physical units (e.g. source extent in arcsec, arcmin, or
beamsizes). On the other hand, a cross-identification tool needs uniform input:
at least a source name, position, and flux density at the measured frequency.
The uniformisation of the 20 SPECFIND V1.0 entry catalogues was done by hand. 
For a significant increase of independent radio cross-identifications,
an input of sources from more than a hundred radio catalogues is needed. This goal could only
be attained by taking advantage of Virtual Observatory (VO) capabilities, which are described in 
Sect.~\ref{sec:uniformisation}.
The new release of the SPECFIND catalogue is presented in Sect.~\ref{sec:specfindv2}.

\section{Table Uniformisation using VO tools \label{sec:uniformisation}}

The Virtual Observatory offers (i) the standards for an efficient table uniformisation
and (ii) the infrastructure to make new tools available to the astronomical community.
The aim of this work is two-fold: 1) discovery of available resources, i.e. radio
catalogues, in the VO world and 2) extraction and homogenisation of the relevant information from these
resources.

Within the framework of the European VO-TECH Design Study we have developed
three VO tools at CDS: (i) TABFIND: a tool to search for useful radio catalogues in the Virtual
Observatory, (ii) TABUNIF: a tool to extract relevant information from these catalogues
and to uniformize the catalogue information, and (iii) CAMEA: a tool to characterize the data,
i.e. to include additional metadata necessary for the full usage of the data in the VO,
as required in the VO ``Characterization'' data model (Louys et al. 2008).
For example, the angular resolution of the observations is not provided as a standard parameter
in the VizieR catalogue description.
During the development the tools were kept as general as possible. 
They can thus be used in other astronomical contexts. We also plan to include the
extended radio catalogue description gathered by the characterization tool into VizieR.
These tools together with the associated manuals are available at http://eurovotech.org/twiki/bin/view/VOTech.

\subsection{Registry query tool TABFIND \label{sec:tabfind}}

This tool identifies VO resources based on Unified Content Descriptors (UCDs).
The UCDs are a controlled vocabulary defined by the VO to describe
astronomical quantities (Derri\`ere et al. 2004).
TABFIND is written in Java and uses XMLDB API
\footnote{http://xmldb-org.sourceforge.net/xapi/} to get data from the VO registry of resources.
The latter is a kind of telephone book where all web servers are listed that comply 
with VO standards.
The user specifies a required set of UCDs. TABFIND searches the VO registry for
all catalogues whose descriptions contain these UCDs.
For example, in our project the minimum set of parameters needed for the
radio cross-identification are source coordinates and a radio flux.

The result of the query is a list of relevant radio catalogues. The catalogues can then be sorted into useful and
not useful catalogues by displaying the catalogue descriptions.
A workspace permits to save and restore all actions performed 
on the catalogues. At the end, a final list of relevant catalogues is established.


\subsection{Data homogenisation tool TABUNIF \label{sec:tabunif}}

The relevant catalogues obtained from TABFIND can be directly loaded into the data homogenization tool.
TABUNIF creates homogenized data from a heterogeneous set of
catalogues. It is written in Java and works on XML tables.
In a first step the user specifies a set of columns for the output table which can
be based on the list of UCDs\footnote{http://www.ivoa.net/Documents/latest/UCDlist.html}. In our case we defined
the output columns according to the needs of the SPECFIND cross-identification tool
(Vollmer et al. 2005a).
In a second step the tool generates an interface where a column of the entry
radio catalogue is assigned to a user specified output column.


The user is free to change the input column that he/she wants to assign to
an output column. It is also possible to assign an arithmetic combination
of different input columns or conditions on input columns to an output column.
As a result the tool generates an ASCII output table for each input radio catalogue. 
The ASCII output tables can be directly used by the SPECFIND cross-identification tool.
Alternatively, the output table can be produced in the VO-compliant VOTable 
format\footnote{The VOTable format is an XML standard for the interchange of data represented as a set of tables; 
http://www.ivoa.net/Documents/latest/VOT.html} for future usage by VO tools.


\subsection{Characterisation tool CAMEA \label{sec:camea}}

We realized that the description of the radio catalogues in VizieR does not contain all
necessary information for the cross-identification.
Basic information such as the identity of the instrument, frequency, resolution, and
observation dates are not included in the catalogue metadata.
We therefore decided to develop a third VO tool which permits to
specify this missing information. More generally, it will permit to
create a full description of a VO resource based on the VO Characterization
data model (Louys et al. 2008). 
In the future, CAMEA will help to provide the input for the data homogenization tool TABUNIF described above.
In this way we plan to complement the VizieR metadata of radio catalogues by adding the frequency, angular 
resolution, observation dates, etc.


\section{SPECFIND V2.0 \label{sec:specfindv2}}

\begin{figure}
\begin{center}
        \resizebox{\hsize}{!}{\includegraphics{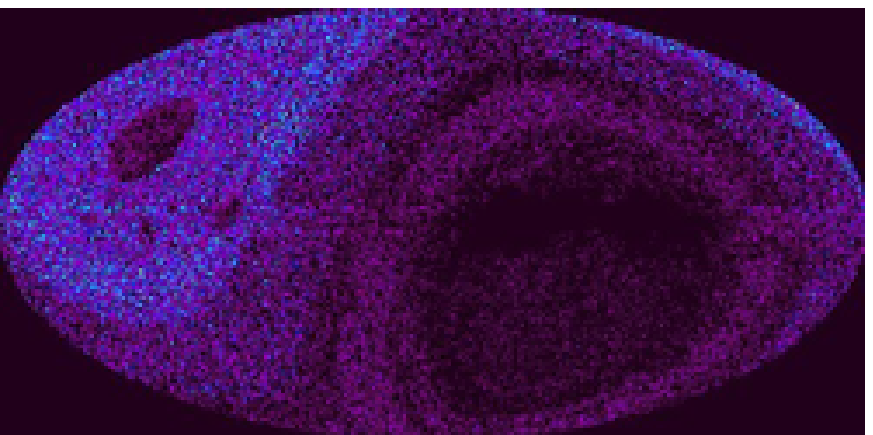}}
	\resizebox{\hsize}{!}{\includegraphics{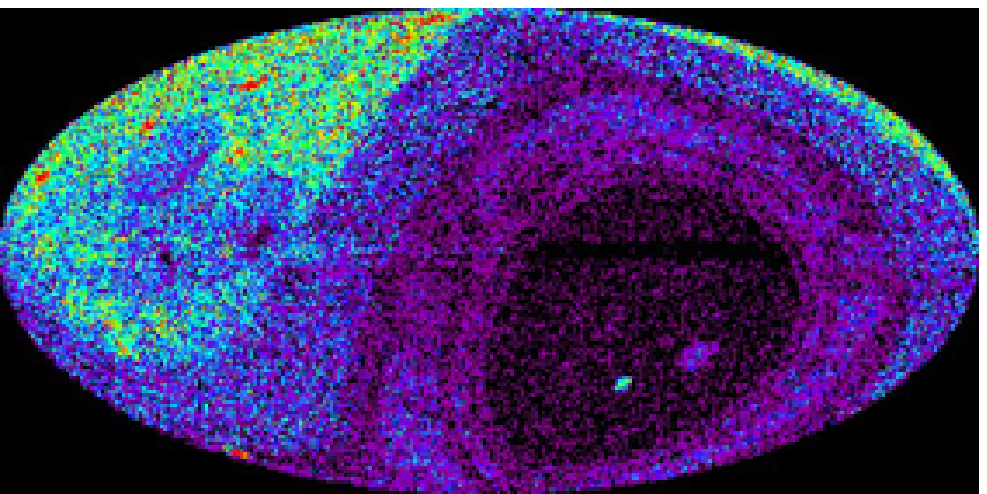}}
        \caption{Sky coverage of radio sources. In both images black/blue/red
	  corresponds to 0/27/64 objects per pixel. Upper panel: from the first release;
	lower panel: from the SPECFIND V2.0 catalogue.
        } \label{fig:skyco}
\end{center}
\end{figure}

The use of the registry query tool and data homogenization tool enabled us
to include 97 radio catalogues and 117 tables (a catalogue corresponds to one reference
and can contained multiple tables) from VizieR (Ochsenbein et al. 2000)
into the SPECFIND radio cross-identification tool (20 catalogues from
Vollmer et al. 2005a (Table~\ref{tab:entries1}) and the catalogues listed in Table~\ref{tab:entries}). 
The number of sources from these catalogues is $3.76 \times 10^6$ giving rise
to 107488 cross-identified objects, i.e. objects with at least 3 flux densities observed
at 3 independent frequencies.
Compared to the first release of the SPECFIND V1.0 catalogue this is an increase of 
available radio sources by $\sim 8$\,\%. This relatively small increase is due to the fact that
the number of catalogues with a given number of radio sources increases rapidly with decreasing number 
of radio sources contained in the catalogue.
However, the smaller catalogues often provide the missing third flux density to 
establish a radio spectrum. For example, in the northern hemisphere there is a multitude of radio 
objects with available NVSS (1.4~GHz;  Condon et al. 1998) and WENSS 
(325~MHz; Rengelink et al. 1997) flux densities. The surveys at higher 
frequencies, which were included in SPECFIND V1.0, are rather shallow and thus did not detect
the majority of the sources. Observations at high frequencies underlying small catalogues
are almost always more sensitive than observations which give rise to large catalogues, with the
drawback that they are made within small areas on the sky. This is the reason why a modest
increase of source available for the cross-identification ($\sim 8$\,\%) leads to
a significant increase of cross-identified radio objects of $\sim 60$\,\%.
The source coverage of the first and the second release of the SPECFIND catalogues are
shown in Fig.~\ref{fig:skyco}. 

The SPECFIND V2.0 catalogue is available via Vizier at CDS. It has the same data structure
as SPECFIND V1.0. Each radio source represents one line of the catalogue. The radio sources
from one physical object are linked via a common sequence number. For each radio source 
SPECFIND V2.0 gives a flag for extended/confused/complex sources (based on the NVSS), 
the source name, coordinates, flux density, error of the flux density as used in
SPECFIND, number of sources with the same sequence number, slope and abcissa of the radio spectrum,
positional difference to the NVSS source which is part of the spectrum, and the difference between
the interpolated 20cm and the NVSS flux density. In addition, we provide a link to the
plot of the radio spectrum and a link to the Aladin applet in which the NVSS/DSS images and the
positions of the SPECFIND V2.0 radio sources together with the beam sizes of the different
observations are displayed. Moreover, we provide access to the radio sources that are cross-identified
only with respect to their position (overlapping beams or extents), but do not fit the radio spectrum.

\begin{table*}	
      \caption{SPECFIND V1.0 catalogue entries.}
         \label{tab:entries1}
         \begin{tabular}{lcrrrrcl}
catalog & I/S$^{1}$ & frequency & resolution & $S_{\rm min}$ & number of & percentage$^{2}$ &  Reference \\
name &  & (MHz) & (arcmin) & (mJy) & sources &  & \\
\hline
JVAS & I &  8400 &  $5.5\,10^{-3}$   & 30 & 2246 & 72 &  Patnaik et al. (1992)  \\
 & & & & & & & Browne et al. (1998) \\
 & & & & & & & Wilkinson et al. (1998) \\
GB6  & S &  4850 &  3.5      &  18 & 75162  & 59 & Gregory et al. (1996) \\
87GB & S &  4850 &  3.5      &  25 & 54579 & 64 & Gregory \& Condon (1991) \\
BWE  & S &  4850 &  3.5      &  25 & 53522  & 61 & Becker et al. (1991) \\
PMN  & S &  4850 &  3.5      &  20 & 50814  & 30 & Wright et al. (1994; 1996) \\
 & & & & & & & Griffith et al. (1994; 1995) \\
MITG & S &  4850 &  2.8      &  40 & 24180 & 52 & Bennett et al. (1986) \\
 & & & & & & & Langston et al. (1990) \\
 & & & & & & & Griffith et al. (1990; 1991) \\
PKS  & S &  2700 &  8.0      &  50 & 8264  & 65 & Otrupcek \& Wright (1991) \\
F3R & S &  2700 &  4.3      &  40 & 6495 & 61 & F\"{u}rst et al. (1990)  \\
FIRST & I & 1400 &  0.083  &  1 & 811117 & 1.5 & White et al. (1998) \\
NVSS  & I & 1400 &  0.75     &  2 & 1773484 & 3.6 & Condon et al. (1998) \\
WB   &  S & 1400 & 10.       & 100 &  31524  & 60  & White \& Becker (1992) \\
SUMSS & I &  843  &  0.75     &  8 & 134870 & 1.8 & Mauch et al. (2003)  \\
B2  & I &  408  &  8.0       & 250 &  9929 & 72 & Colla et al. (1970; 1972; 1973)  \\
 & & & & & & & Fanti et al. (1974) \\
B3 & I & 408  &  5.0      &  100 & 13340 & 66 & Ficarra et al. (1985) \\
MRC & I & 408 &  3.0      &  700 & 12141 & 73 &  Douglas et al. (1996)   \\
TXS  & I &  365  &  0.1      &  250 & 66841 & 57 & Large et al. (1991) \\
WISH & I & 325  &  0.9      &  10 & 90357 & 8.4 & de Breuck et al. (2002) \\
WENSS & I & 325  &  0.9      &  18 & 229420 & 17 & Rengelink et al. (1997) \\
MIYUN & I & 232  &  3.8      &  100 & 34426 & 40 & Zhang et al. (1997) \\
4C  & I &  178  &  11.5     &  2000 & 4844 & 53 & Pilkington \& Scott (1965)  \\
 & & & & & & & Gower et al. (1967) \\
3CR & I & 178  &  6.0       &  5000 & 327 & 31 & Bennet (1962) \\
3C   & I & 159  &  10.0      &  7000 & 470 & 3.4 & Edge et al. (1959) \\ 
\end{tabular}
\begin{list}{}{}
\item
$^{1}$: S: single dish; I: interferometer
\item
$^{2}$: Percentage of sources with identified spectrum. Only unambiguous source names could be counted.
\end{list}
\end{table*}

\begin{table*}	
      \caption{Additional SPECFIND V2.0 catalogue entries.}
         \label{tab:entries}
         \begin{tabular}{lcrrrrcl}
catalog & I/S$^{1}$ & frequency & resolution & $S_{\rm min}$ & number of & percentage$^{2}$ &  Reference \\
name &  & (MHz) & (arcmin) & (mJy) & sources &  & \\
\hline
    NEK  &   I &     31 &  12.0  & 2000   &   703 &   4  &  Kassim (1988) \\
    8C   &   I  &    38  &  4.5   & 700   &  5859 &  66  &  Hales et al. (1995) \\
    CRJ2004& I  &    74  &  0.42  & 150   &   949 &  63  &  Cohen et al. (2004) \\
    VLSS  &  I &     74  &  1.33  & 400   & 68308 &  66 &  Cohen et al. (2007) \\
    TRC2006b&I &     74 &   0.5   &   0.3 &   725 &   0.2  &  Tasse et al. (2006) \\
    Cula  &  I &     80  &  3.7  & 2000   &  2173 &  49  &  Slee (1995) \\
    6C  &    I  &   151  &  4.2   & 300   & 27666 &  84  & Baldwin et al. (1985) \\ 
     & & & & & & & Hales et al. (1988/80/91/93) \\
    7C  &    I  &   151  &  1.17  & 200   & 43683 &  77  &  Hales et al. (2007) \\
    Culb  &  I &    160  &  1.85 & 1000   &  2042 &  66  &  Slee (1995) \\
    TRB2007a&I  &   240  &  0.25  &   6   &   466 &  63  &  Tasse et al. (2007) \\
    TRC2006a&I &    325  &  0.12  &   2.5 &   843 &  47  &  Tasse et al. (2006) \\
    WSTB90&  I  &   327  &  1.5   &   3   &   407 &      &  Oort et al. (1988) \\
    W93a  &  I &    327  &  1.0   &   1   &  4157 &  32  &  Wieringa (1993) \\
    WSTB2 &  I &    327  &  1.5  &   10   &   309 &      &  Righetti et al. (1988) \\
    WSRTGP & I &    327 &   1.0   &   5   &  3984 &  14  &  Taylor et al. (1996) \\
    PSRa  &    &    400 &   1.0   &   0.1 &   561 &   0.4  &  Taylor et al. (1993) \\
    32P   &  I  &   408  &  4.0   &  30   &   494 &  61  &  Leahy \& Roger (1996) \\
    5C5  &   I &    408  &  1.5   &  10   &   214 &  60  &  Pearson (1975) \\
    5C6   &  I &    408  &  1.5   &   6   &   267 &  68  &  Pearson (1978) \\
    5C7   &  I  &   408  &  1.5   &  10   &   235 &  41  &  Pearson (1978) \\
    5C12  &  I &    408  &  1.5  &    2   &   308 &  70  &  Benn et al. (1982) \\
    5C13 &   I &    408  &  1.5   &  12   &   238 &  60  &  Benn (1995) \\
    51P  &   I &    408  &  4.0   &  80   &   383 &  31  &  Green \& Riley (1995) \\
    5C12a &  I &    408  &  1.33 &    5   &   680 &  49  &  Benn \& Kenderdine (1991) \\
    DRAOP &  I &    408  &  3.5   &       &   915 &  62  &  Landecker \& Caswell (1983)\\
          &    &         &        &       &       &      &  VizieR VIII/55$^{3}$\\
    CGPSEa & I &    408  &  3.0   &  10   &   140 &  24  &  Kerton et al. (2007) \\
    PSRb  &   &     600  &  1.0   &   0.4 &   352 &   0.4  &  Taylor et al. (1993) \\
    W93b  &  I &    608  &  0.5   &   1   &  1693 &  62  &  Wieringa (1993) \\
    TRB2007b&I  &   610  &  0.11  &   1  &   1037 &  35  &  Tasse et al. (2007) \\
    FLSGMRT &I  &   610  &  0.1   &   0.1 &  3944 &   1  &  Garn et al. (2007) \\
    NAIC  &  S &    611  & 12.6   & 350   &  3122 &  45  &  Durdin et al. (1975) \\
    MOST  &  I &    843  &  0.733 &  40   &   348 &  59  &  Jones \& McAdam (1992) \\
    MGPS2  & I  &   843  &  0.75  &  10  &  48850  &  3  &  Murphy et al. (2007) \\
    ATESP  & I &   1400  &  0.23  &   0.3 &  3370 &   1  &  Wieringa \& Ekers (2000) \\
    PSRc &     &   1400  &  1.0   &   0.1 &   445 &   0.4  &  Taylor et al. (1993) \\
    OWH82 &  S &   1400  & 10.0   & 100   &   487 &  63  &  Owen et al. (1982) \\
    GPSR  &  I &   1400  &  0.083 &   5   &  1992 &  11  &  Zoonematkermani et al. (1990) \\
    LO95  &  I  &  1400  &  0.25  &  10   &   375 &      &  Ledlow \& Owen (1995) \\
    PDF   &  I &   1400  &  0.15  &   0.1 &  1079 &   0.1  &  Hopkins et al. (1998) \\
    VIRMOS&  I &   1400  &  0.1  &    0.1 &  1103 &   4  &  Bondi et al. (2003) \\
    FHW95a & S &   1400  & 15.2   &  40   &   192 &  45  &  Filipovic et al. (1995) \\
    WST32 &  I &   1400  &  1.2  &   10   &   215 &  29  &  Fanti et al. (1981) \\
    37W   &  I &   1400  &  0.6  &    1   &    53 &  28  &  Walterbos et al. (1985) \\
    ELAISR & I &   1400  &  0.25  &   0.1 &   965 &   1  &  Ciliegi et al. (1999) \\
    MGC2004& I &   1400  &  0.23  &   0.03&  1048 &   1  &  Morganti et al. (2004) \\
    WBH2005 &I &   1400  &  0.1   &   2  &   6919  &  4  &  White et al. (2005) \\
    RRF1  &  S  &  1410  &  9.4   & 100   &   884 &      &  Reich et al. (1990) \\
    RRF   &  S &   1410  &  9.4   &  80   &  1830 &      &  Reich et al. (1997) \\
    QC93  &  S &   1410 &  10.0  &  400   &   171 &  57  &  Quiniento \& Cersosimo (1993) \\
    CCH85  & I &   1411 &   0.33 &    5   &   208 &   9  &  Coleman et al. (1985) \\
    WSTB &   I &   1412 &   0.385 &   1   &   536 &      &  Windhorst et al. (1984) \\
    WSTB1 &  I &   1412  &  0.385 &   1   &   359 &      &  Oort (1987) \\
\end{tabular}
\begin{list}{}{}
\item
$^{1}$: S: single dish; I: interferometer
\item
$^{2}$: Percentage of sources with identified spectrum. Only unambiguous source names could be counted.
\item
$^{3}$: This catalogue is a compilation of tables in 27 articles. Landecker \& Caswell (1983) is the 
first reference.
\end{list}
\end{table*}

\addtocounter{table}{-1}
\begin{table*}	
      \caption{Additional SPECFIND V2.0 catalogue entries (continued).}
         \begin{tabular}{lcrrrrcl}
catalog & I/S$^{1}$ & frequency & resolution & $S_{\rm min}$ & number of & percentage$^{2}$ &  Reference \\
name &  & (MHz) & (arcmin) & (mJy) & sources &  & \\
\hline
    33P    & I  &  1420  &  1.0   &   3   &   255 &  36  &  Leahy \& Roger (1996) \\
    FBR2002& I &   1420  &  1.633 &   3   &   534 &  43  &  Filipovic et al. (2002) \\
    CGPSEb & I  &  1420 &   1.0   &   3  &    140  & 27  &  Kerton et al. (2007) \\
    RLM94  & I  &  1465  &  0.075 &   5   &   725 &  42  &  Roettgering et al. (1994) \\
    P82\_1 &  I  &  1465  &  0.43  & 200   &   404 &  71  &  Perley (1982) \\ 
    NEP  &   I  &  1490  &  0.333 &   1   &  2435 &  16  &  Kollgaard et al. (1994) \\
    DC78 &   S  &  2370  &  2.7   &   1   &   858 &   29  &  Dressel \& Condon (1978) \\
    FBR2002a&I &   2370  &  0.667 &   1   &   697 &  33  &  Filipovic et al. (2002) \\
    FHW95b & S &   2450  &  8.85 &   30   &   334 &  54  &  Filipovic et al. (1995) \\
    FORa  &  S &   2695  &  4.78  &  20   &   221 &  81  &  Forkert \& Altschuler (1987) \\
    RFS  &   S  &  2695  &  4.9   &  30   &  1212 &  27  &  Reich et al. (1984) \\
    RGB11 &  S  &  2700  &  4.3   &  50   &   697 &  50  &  Reich et al. (2000) \\
    PBD2003& S  &  2700  &  8.0   & 100   &  1432 &  9  &  Paladini et al. (2003) \\
    KMP90 &  S  &  4730  &  2.8   &  15   &   752 &  85  &  Kulkarni et al. (1990) \\
    FORb  &  S &   4750  &  2.71  &  25   &   227 &  85  &  Forkert \& Altschuler (1987) \\
    FHW95c & S &   4750  &  4.8  &   15   &   368 &  53  &  Filipovic et al. (1995) \\
    NAICGB & S &   4775  &  2.8   &   8  &   2453 &  65  &  Lawrence et al. (1983) \\
    A86    & S &   4760  &  2.8  &   15   &   882 &  74  &  Altschuler (1986) \\
    FBR2002b&I &   4800 &   0.5   &   1   &    75 &  40  &  Filipovic et al. (2002) \\
    GDP   &  I &   4850  &  0.5  &    0.3 &   253 &   8  &  Gregorini et al. (1994) \\
    CAB95 &  S  &  4850  &  3.5   &  25   &   351 &  76  &  Condon et al. (1995) \\
    JRB99  & I &   4860  &  0.03 &    1   &   298 &  28  &  Jackson et al. (1999) \\
    FPD2001a&I &   4860  &  6.7e-3&   1   &   213 &  36  &  Fanti et al. (2001) \\
    ADP79  & S &   4875  &  2.6   & 100   &   569 &  19  &  Altenhoff et al. (1979) \\
    RGB   &  I  &  4885  &  0.067 &  13   &  1861 &  56  &  Laurent-Muehleisen et al. (1997) \\
    P82\_2  & I  &  4885  &  0.1   & 200   &   404 &  82  &  Perley (1982) \\ 
    KR    &  I &   4890  &  0.07 &   10   &   195 &  41  &  Fich (1986) \\
    GPSR5 &  I  &  4900  &  0.067 &   1   &  1286 &   3  &  Becker et al. (1994) \\
    Slee  &  I &   4900  &  0.67  &  0.2  &   177 &  21  &  Slee et al. (1998) \\
    HCS79 &    &   5000  &  0.54  & 240   &   702 &   3  &  Haynes et al. (1979) \\
    RGB6  &  S  &  5000  &  2.4   &  40   &   729 &  56  &  Reich et al. (2000) \\
    GPA2  &  S  &  8350  &  9.7   & 900   &   555 &   6  &  Langston et al. (2000) \\
    CLASS &  I &   8400  &  5.5e-3&   0.1 & 21486 &  39  &  Myers et al. (2003) \\
    FPD2001b&I &   8460  &  3.3e-3&   0.2&    199 &  33  &  Fanti et al. (2001) \\
    FHW95d&  S &   8550  &  2.7  &   20   &   205 &  47  &  Filipovic et al. (1995) \\
    FBR2002c&I &   8640  &  0.27  &   1   &    54 &  41  &  Filipovic et al. (2002) \\
    NKB95 &  S &  10550  &  1.15 &    3   &   202 &  57  &  Niklas et al. (1995) \\
    B3VLA &  S  & 10600  &  1.15  &  10   &   981 &  80  &  Gregorini et al. (1998) \\    
    RGB2  &  S  & 10700  &  1.2   &  30   &   698 &  53  &  Reich et al. (2000) \\
    GPA1 &   S &  14350  &  6.6   &2000   &   365 &   7  &  Langston et al. (2000) \\
    9C    &  I &  15000  &  0.42  &  25   &   242 &  71  &  Waldram et al. (2003) \\
\end{tabular}
\begin{list}{}{}
\item
$^{1}$: S: single dish; I: interferometer
\item
$^{2}$: Percentage of sources with identified spectrum. Only unambiguous source names could be counted.
\end{list}
\end{table*}

\begin{figure}
\begin{center}
        \resizebox{\hsize}{!}{\includegraphics{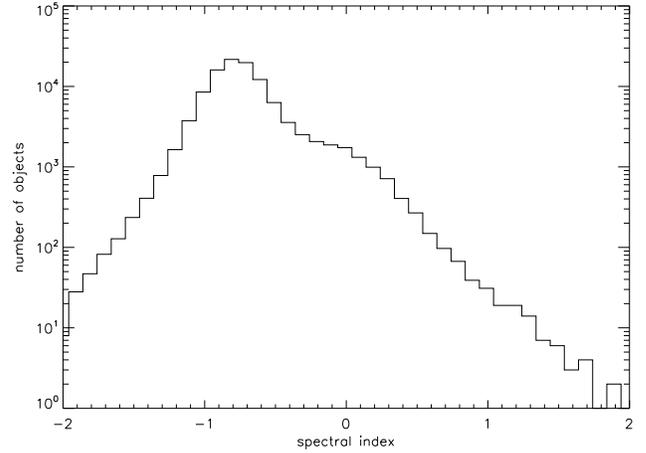}}
        \caption{Number distribution of spectral indices.
        } \label{fig:numbersi}
\end{center}
\end{figure}

The distribution of the spectral indices\footnote{The
spectral index $\alpha$ is defined by $S_{\nu} \propto \nu^{\alpha}$.} is shown in Fig.~\ref{fig:numbersi}.
The distribution peaks at $\alpha \sim -0.9$, which is consistent with the result
from the first release (see also Zhang et al. 2003).
As in SPECFIND V1.0, there is a wing towards positive spectral indices, which is most probably caused 
by the flattening of the spectrum at low frequencies due to synchrotron self-absorption and
at high frequencies due to the emission of thermal electrons.

\begin{figure}
\begin{center}
        \resizebox{\hsize}{!}{\includegraphics{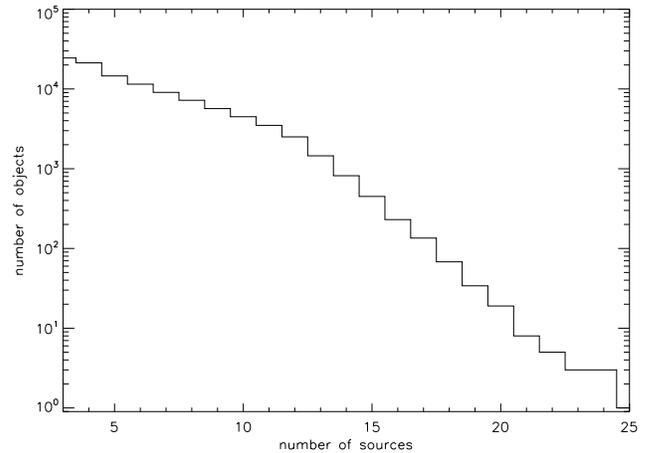}}
        \caption{The number of sources as a function of the number of independent points in the
	  radio spectrum.
        } \label{fig:numberofs}
\end{center}
\end{figure}

The number of objects as a function of the number of catalogued sources contained in an object
is shown in Fig.~\ref{fig:numberofs}.
The shape of the distribution is a broken power law with a break at about 12 sources per object.
The maximum number of catalogued sources in a physical object is 30.

\begin{figure}
\begin{center}
        \resizebox{\hsize}{!}{\includegraphics{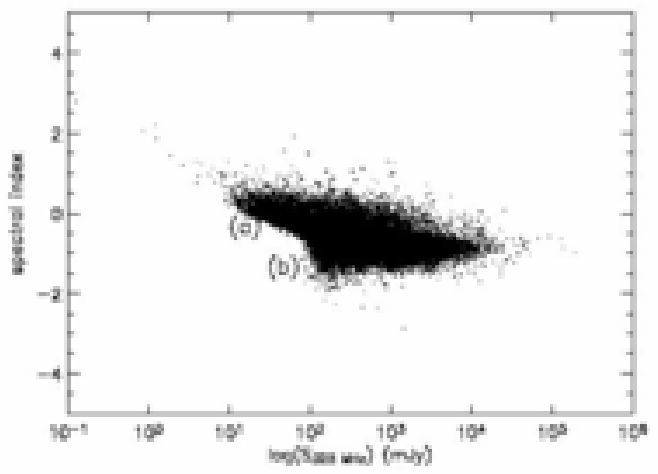}}
	\resizebox{\hsize}{!}{\includegraphics{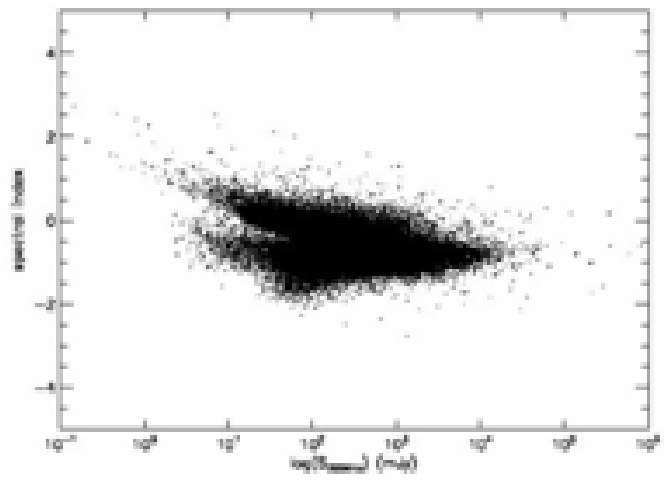}}
        \caption{Spectral index as a function of 325~MHz flux density. Upper panel: SPECFIND V1.0
	  (Vollmer et al. 2005a). Lower panel: SPECFIND V2.0.
        } \label{fig:sidist}
\end{center}
\end{figure}

The distributions of the spectral indices as a function of the measured or interpolated flux
density at 325~MHz from SPECFIND V1.0 and V2.0 are shown in Fig.~\ref{fig:sidist}.
In SPECFIND V1.0 the straight, almost horizontal edge of the distribution in the left part of the plot 
(marked as (a) in the upper panel of Fig.~\ref{fig:sidist}) is due to a selection effect.
For these low flux density sources with a steep spectrum, SPECFIND found a source at 20~cm (NVSS)
and 50~cm (WENSS), but none at 6~cm, where the sensitivity
of the surveys ($\sim 20$~mJy) is insufficient.
The inclusion of the new radio catalogues at 4850~MHz into SPECFIND V2.0 improved this situation only mildly. 

The vertical edge in the lower left part of the plot (marked as (b) in the upper panel of
Fig.~\ref{fig:sidist}) is mainly due to the limiting flux density of the B3 survey. 
Here, the inclusion of a significant number of new sources at low frequencies ($\nu < 1$~GHz,
mainly the 6C and 7C catalogues) leads to a significant increase of objects with 325~MHz flux densities 
smaller than 100~mJy and spectral indices smaller than 0.

\section{Compatibility with SPECFIND V1.0 \label{sec:compatibility}}

The cross-identification of radio sources observed at different frequencies and
with considerably different angular resolutions (see Table~\ref{tab:entries}) is a complex task.
The details of the SPECFIND cross-identification algorithm are described in Vollmer et al.
(2005). In a first step SPECFIND makes a positional cross-identification accounting
for the source extent and resolution of the survey. In a second step the flux densities
and associated errors observed at the same frequency are compared and in a third step, a power law
is fitted to the flux densities at different frequencies.
The cross-identification is done for each catalogued radio source separately
resulting in different spectral indices of sources belonging to the same object.
In crowded fields with a high source density the cross-identification might not be unique and
depends on the weight given to each source in the physical object. The resulting 
degeneracy in the cross-identification is solved by a self-consistency check of all
physical objects found by SPECFIND (for details see Vollmer et al. 2005a).
This procedure insures that a radio source cannot be associated to two different
physical objects. 

Once sources of new catalogues are added to the input of the SPECFIND cross-identification
tool, sources in crowded fields can be redistributed among physical objects, former
objects can disappear and new ones can be created. 
To insure the compatibility between the SPECFIND V1.0 and V2.0 catalogues, we developed
a tool to check the coherence between these catalogues.
This tool searches for (i) radio sources of SPECFIND V1.0 which are not found in V2.0 and (ii)
radio sources of a physical object in SPECFIND V1.0 which are found in different
physical objects in V2.0. The tool displays the V1.0 and V2.0 object lists together with
the V1.0 and V2.0 spectra. As an additional step the user can display the NVSS 20~cm image
within Aladin (Bonnarel et al. 2000) together with the source positions and the beam sizes (angular resolution) 
(Fig.~\ref{fig:source1} -- Fig.~\ref{fig:source14}).
This information allowed us to (i) merge both spectra, or to keep the (ii) the V1.0 or 
(iii) V2.0 spectrum. In this way we visualized and modified about 1000 physical objects
in the SPECFIND V2.0 catalogue.

This procedure also allowed us to detect complicated cases of the radio cross-identification.
In the following we present the 5 major classes of physical objects that a user finds
in the SPECFIND V2.0 catalogue: (i) well-behaved, (ii) extended, (iii) complex, (iv) physical double,
and (v) unphysical double sources. Sources of class (ii)--(v) can be recognized by
a spread of spectral indices $\alpha$ of the sources contained in a physical object which is
larger than the uncertainty due to the flux density errors ($\Delta \alpha > 0.3$; Vollmer et al. 2005b).  
We decided to leave all these sources in the catalogue.
{\it We therefore caution the user against a blind use of the SPECFIND V2.0 catalogue.}
To help the user, we provide flags for sources which 
(i) have at least one neighboring NVSS source within a radius of $2'$ (possible confusion), (ii)
have deconvolved sizes larger than $45''$ in the NVSS catalogue (extended sources), and 
(iii) are marked as complex in the NVSS catalogue.

\subsection{Well-behaved sources}

\begin{figure}
\begin{center}
        \resizebox{\hsize}{!}{\includegraphics{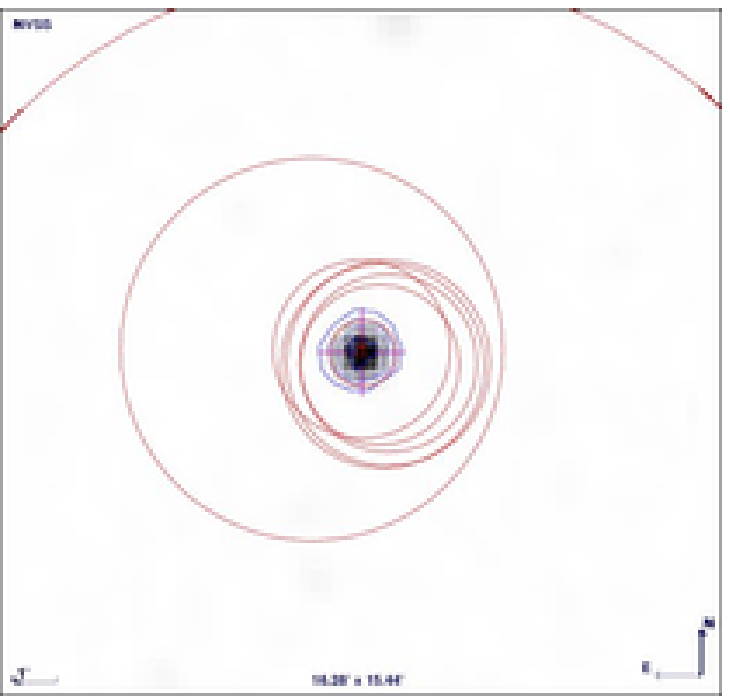}}
	\resizebox{\hsize}{!}{\includegraphics{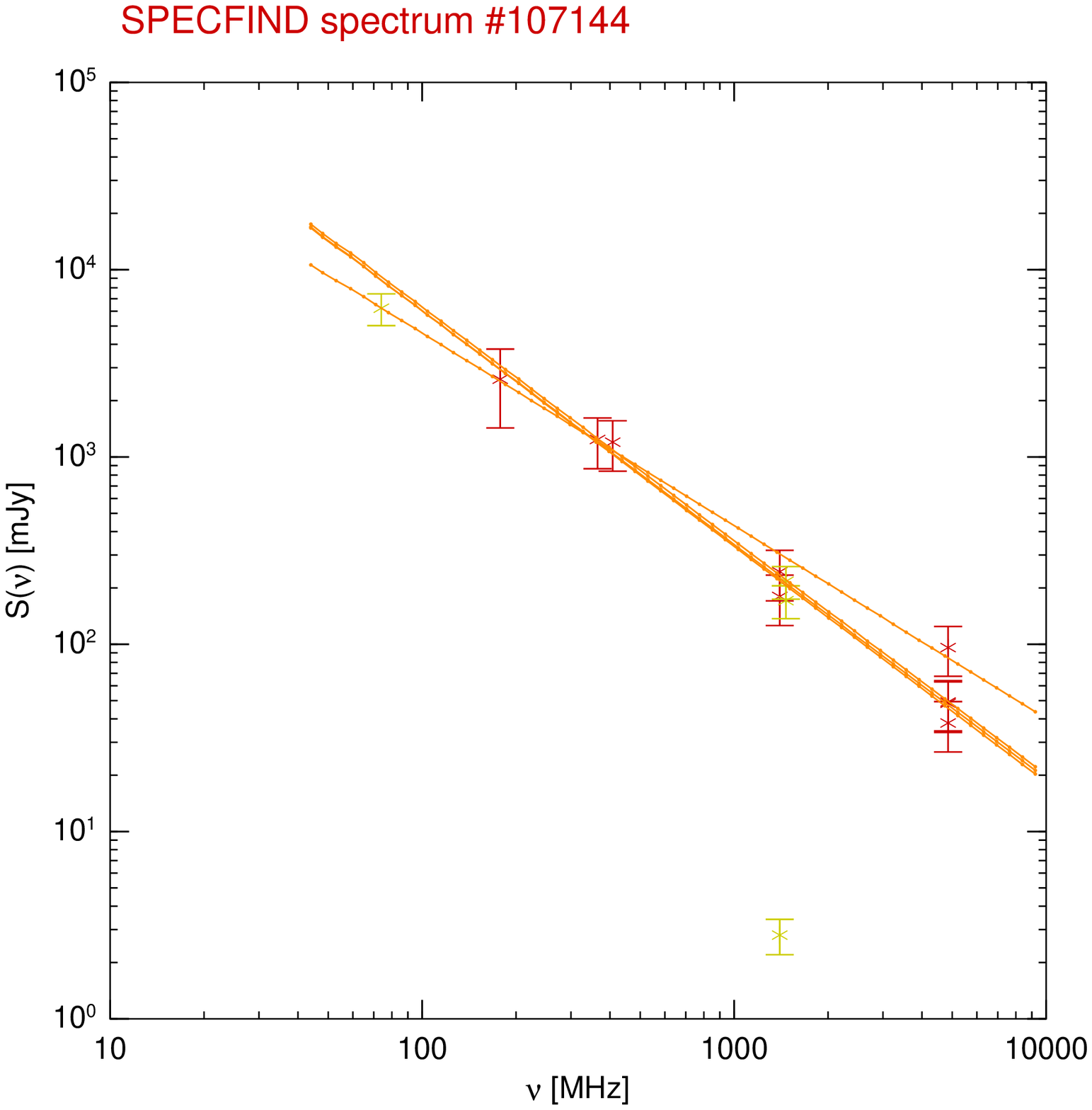}}
        \caption{Well-behaved sources - TXS 2112+158.
	  Upper panel: Aladin view of the NVSS image, the positions of the radio sources
	  and the beamsizes of the radio surveys. Lower panel: 
	  Vizier view of the radio spectrum. Red symbols: Specfind V2.0; green symbols:
	  waste, i.e. source with overlapping beams that do not fit the radio spectrum.
        } \label{fig:source1}
\end{center}
\end{figure}

The vast majority of the SPECFIND V2.0 objects are well-behaved
(Fig.~\ref{fig:source1}), i.e. they are
unresolved or marginally resolved in most of the surveys and exhibit consistent
spectral indices for all sources (the uncertainty of the spectral index is
$\pm 0.3$; Vollmer et al. 2005a). In the case of sources which are only present in
one of the catalogues SPECFIND V1.0 or V2.0, we merged the spectra of all these well-behaved sources. 

\subsection{Extended sources}

\begin{figure}
\begin{center}
        \resizebox{\hsize}{!}{\includegraphics{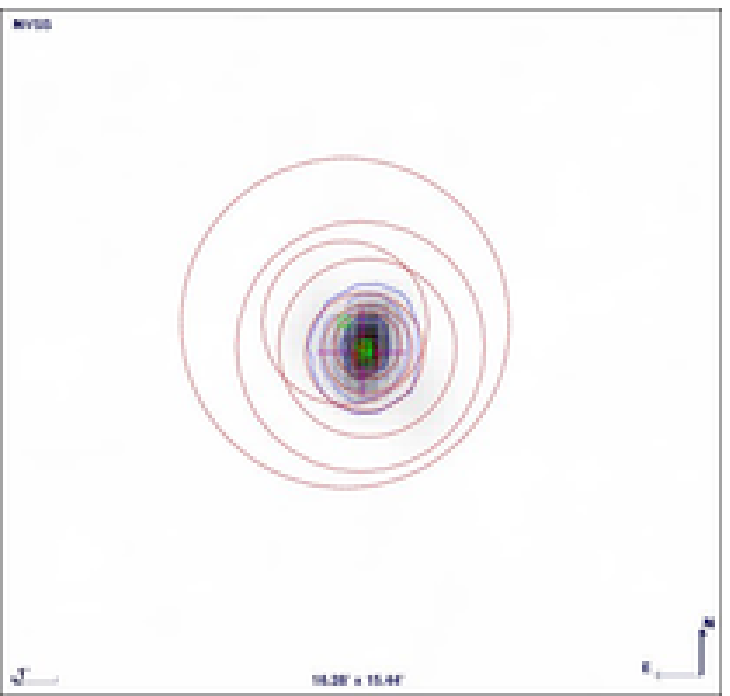}}
	\resizebox{\hsize}{!}{\includegraphics{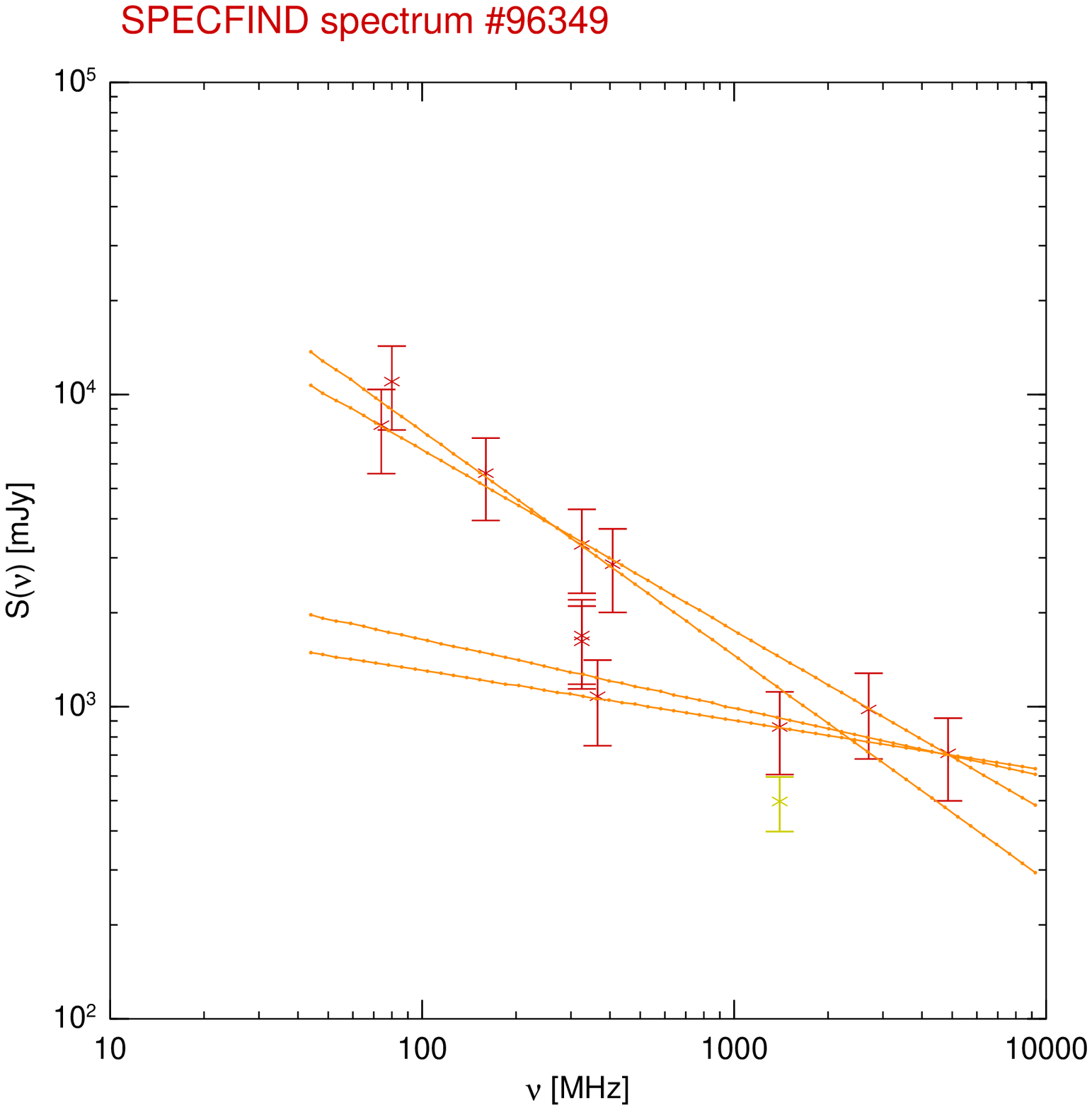}}
        \caption{Extended sources - TXS 2300-189.
	    Upper panel: Aladin view of the NVSS image, the positions of the radio sources
	  and the beamsizes of the radio surveys. Lower panel: 
	 Vizier view of the radio spectrum. Red symbols: Specfind V2.0; green symbols:
	  waste, i.e. source with overlapping beams that do not fit the radio spectrum.
        } \label{fig:source10}
\end{center}
\end{figure}

The second class of sources are those which are extended with respect to the mean
resolution of the input catalogues which is $\sim 1$--$2'$.
An example for such a source is shown in Fig.~\ref{fig:source10}.
At low frequencies these objects have radio fluxes from low resolution surveys in which the source is
unresolved. However, at high frequencies the radio fluxes are provided by high resolution surveys. 
The source is thus resolved leading to a flux density which is smaller than the total
flux density. 
The spectrum of the physical object has two different slopes: one which is fitted to
the flux density of the unresolved sources and one which is fitted to
the flux density of the resolved sources. In these cases the user has to
verify the resolutions of the data points and to make a choice to
which points he or she wants to fit a power law. 
Extended sources thus exhibit different spectral indices ($\Delta \alpha > 0.3$) within
the same frequency range (between 100 and 500~MHz in the example of Fig.~\ref{fig:source10}).

\subsection{Complex sources}

\begin{figure}
\begin{center}
        \resizebox{\hsize}{!}{\includegraphics{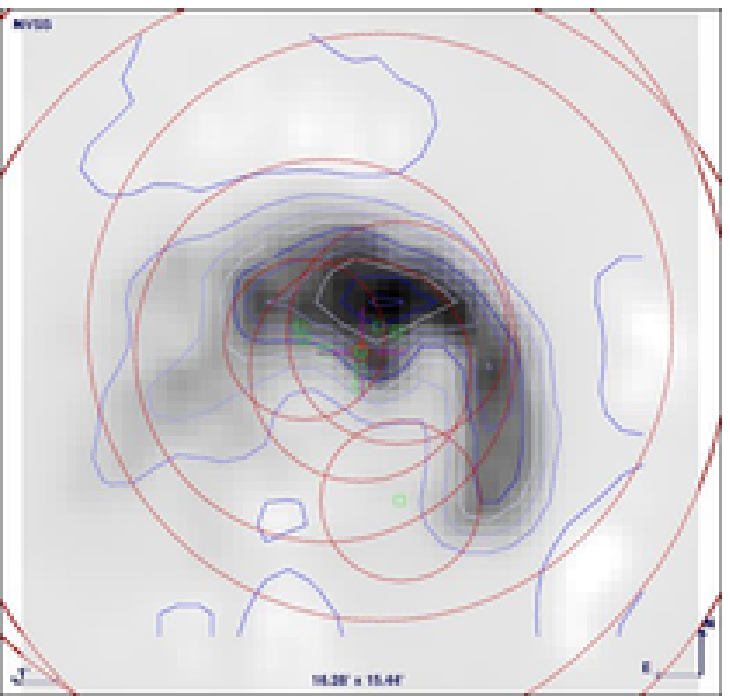}}
	\resizebox{\hsize}{!}{\includegraphics{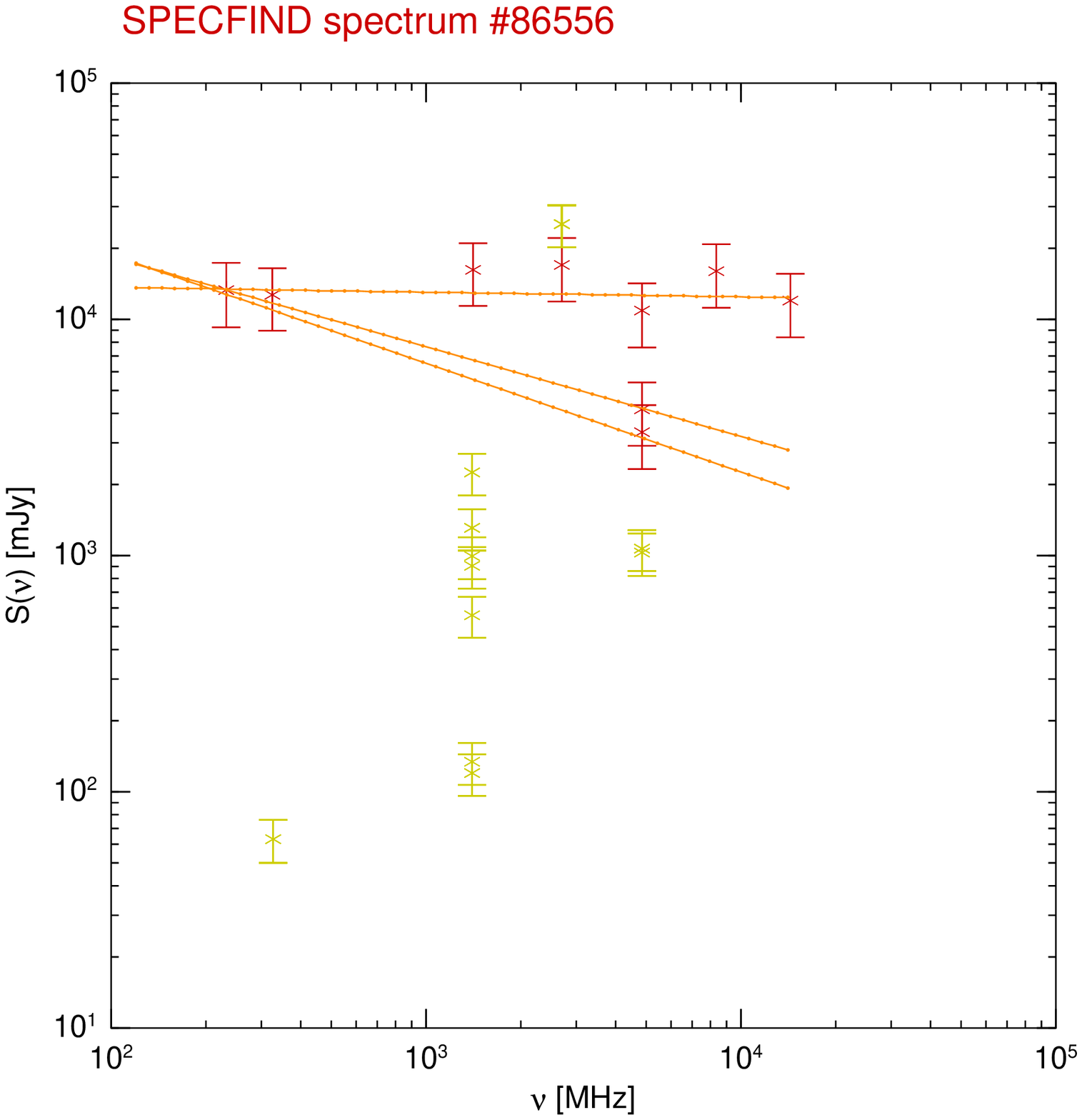}}
        \caption{Complex sources - WN B2040.8+4246.
	 Upper panel: Aladin view of the NVSS image, the positions of the radio sources
	  and the beamsizes of the radio surveys. Lower panel: 
	  Vizier view of the radio spectrum. Red symbols: Specfind V2.0; green symbols:
	  waste, i.e. source with overlapping beams that do not fit the radio spectrum.
        } \label{fig:source11}
\end{center}
\end{figure}

Nearby large Galactic radio sources often display a complex structure (Fig.~\ref{fig:source11}).
In the presence of a sufficient number of observations at different frequencies
the SPECFIND algorithm identifies power laws, but as for the extended sources,
there are multiple spectral indices within the same frequency range (here between
100~MHz and 10~GHz). In the case of such complex sources the user has to carefully
inspect all resolutions and, if necessary, all source extents in the original catalogues in VizieR.

\subsection{Physical double sources}

\begin{figure}
\begin{center}
        \resizebox{\hsize}{!}{\includegraphics{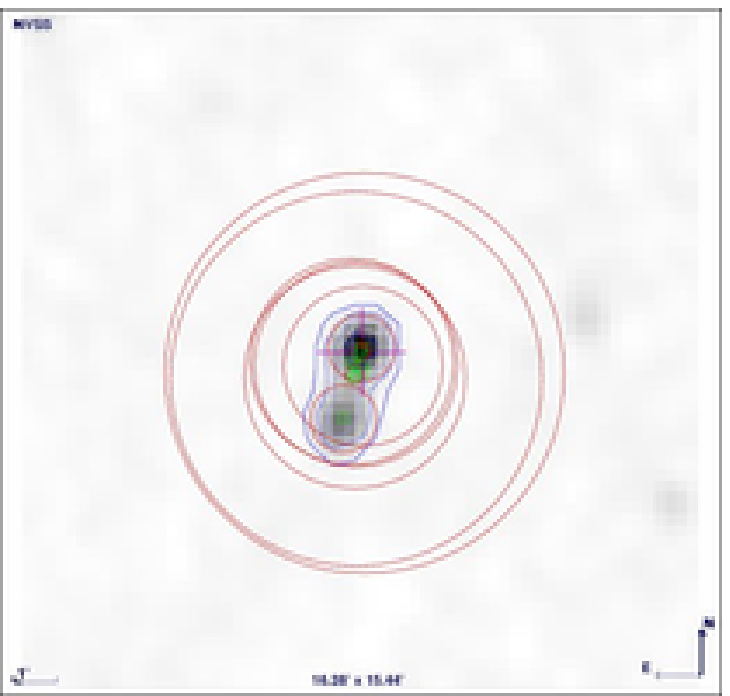}}
	\resizebox{\hsize}{!}{\includegraphics{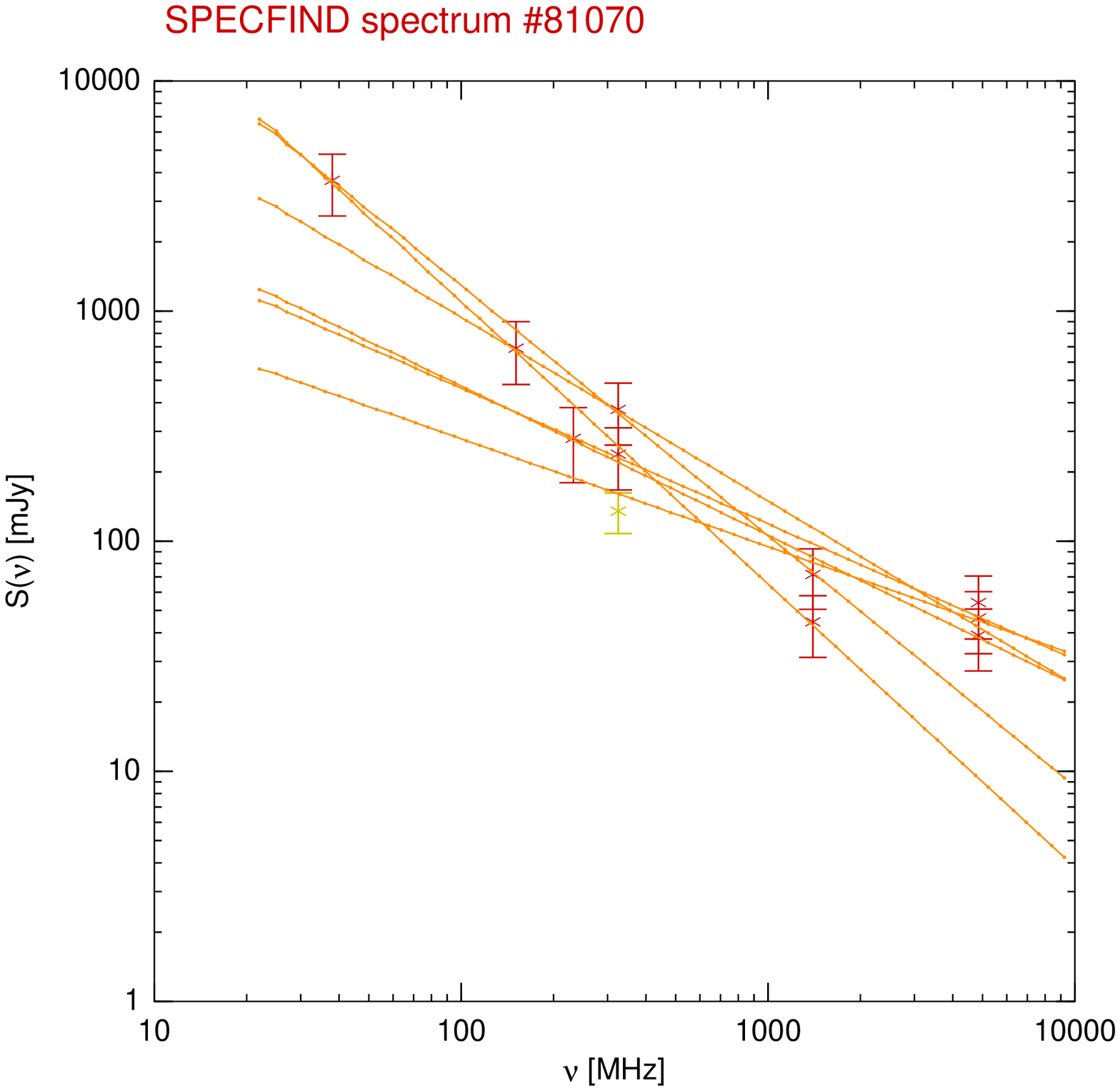}}
        \caption{Possibly physical double sources - WN B1853.1+6226A.
	  Upper panel: Aladin view of the NVSS image, the positions of the radio sources
	  and the beamsizes of the radio surveys. Lower panel: 
	  Vizier view of the radio spectrum. Red symbols: Specfind V2.0; green symbols:
	  waste, i.e. source with overlapping beams that do not fit the radio spectrum.
        } \label{fig:source13}
\end{center}
\end{figure}

Since the typical resolution of the SPECFIND entry surveys is $\sim 2'$ (Table~\ref{tab:entries}),
radio sources which are separated by less than this distance will most frequently
end up in one object in the SPECFIND catalogue (Fig.~\ref{fig:source13}).
There are observations with large beamwidths which comprise the two sources and observations
with higher resolution which resolve the two sources. In a physical double source,
i.e. the radio lobes of an active galactic nucleus, the two sources often differ in
flux density, size, and spectrum. Many double lobe radio sources show an intrinsic
asymmetry in their lobes, probably due to Doppler-boosting and jet inclination.
The different resolutions of the surveys and the detection of different source components leads 
to a dispersion of the spectral indices in  the composite spectrum.

\subsection{Unphysical multiple sources}

\begin{figure}
\begin{center}
        \resizebox{\hsize}{!}{\includegraphics{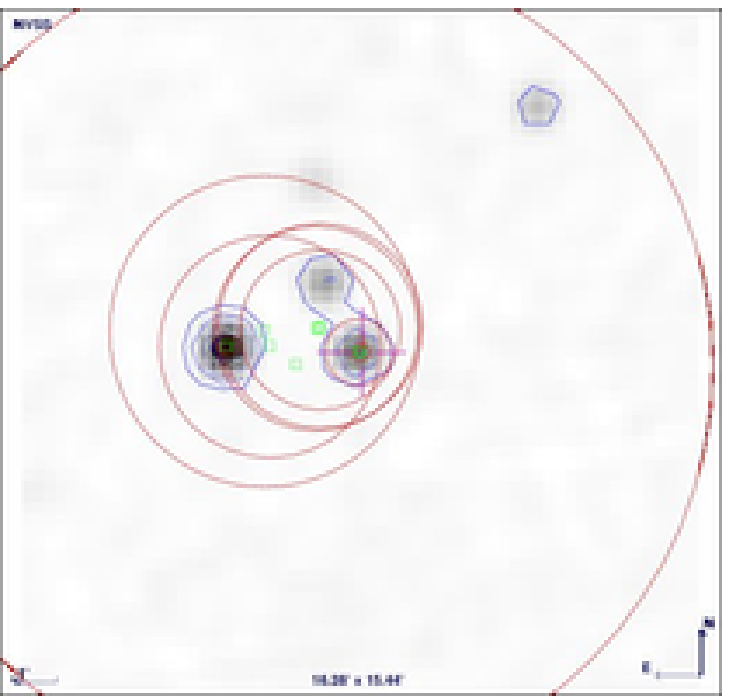}}
	\resizebox{\hsize}{!}{\includegraphics{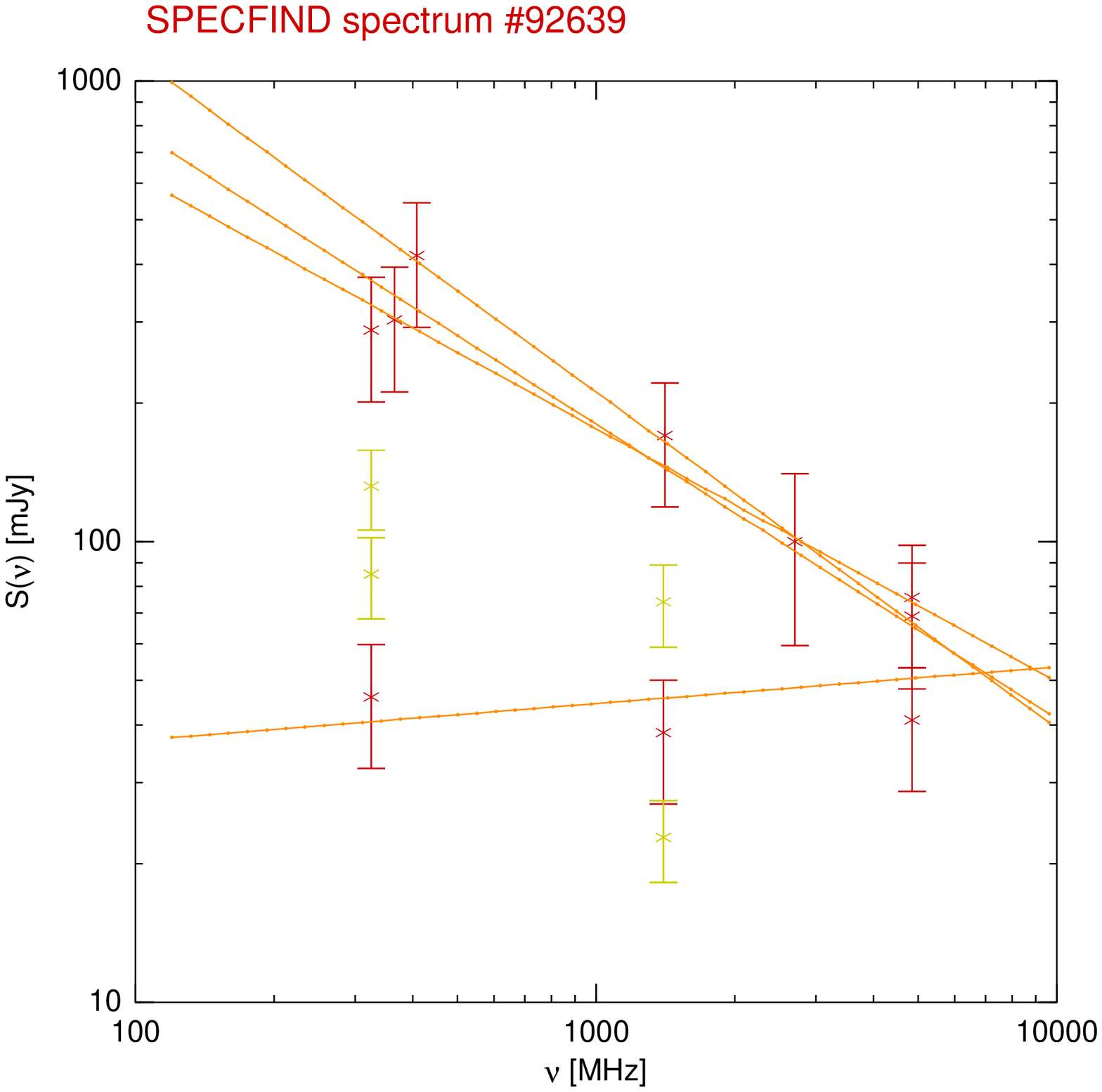}}
        \caption{Unphysical double sources - WN B2228.2+5940A.
	  Upper panel: Aladin view of the NVSS image, the positions of the radio sources
	  and the beamsizes of the radio surveys. Lower panel: 
	  Vizier view of the radio spectrum. Red symbols: Specfind V2.0; green symbols:
	  waste, i.e. source with overlapping beams that do not fit the radio spectrum.
        } \label{fig:source14}
\end{center}
\end{figure}

Other sources which are separated by less than $2'$ are not physically related, i.e.
the sources are only close in projection (Fig.~\ref{fig:source14}).
When these sources exhibit two largely different spectral indices, they can be separated
easily with the help of the radio spectrum.
If these source have about the same spectral index the user has to rely on the NVSS image.

\section{Spectral breaks \label{sec:sibreaks}}

\begin{figure}
\begin{center}
        \resizebox{\hsize}{!}{\includegraphics{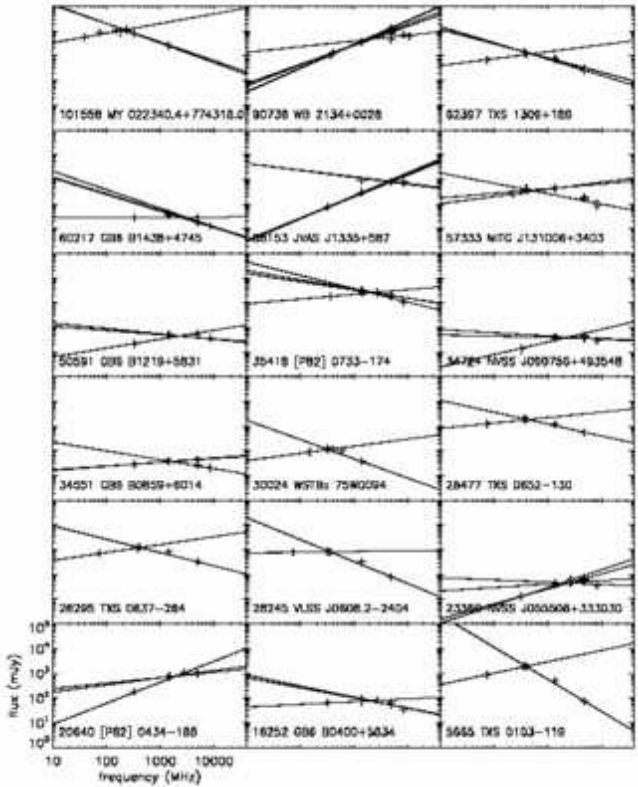}}
        \caption{Radio objects with spectral breaks identified by SPECFIND. 
	  The different lines correspond to fits of different parts of the radio spectrum.
	  The SPECFIND V2.0 
	  sequence number from VizieR and the name of one of the radio sources are marked in each box.
        } \label{fig:sibreaks}
\end{center}
\end{figure}

As described in Vollmer et al. (2005a) the SPECFIND algorithm is able to detect
a spectral break if enough data points at independent frequencies are available.
Due to the increased number of input frequencies in the SPECFIND V2.0 catalogue,
we could identify 18 sources which show a spectral break (Fig.~\ref{fig:sibreaks}).
Since the frequency coverage of the input survey ranges between $\sim 100$~MHz and $\sim 10$~GHz,
the break is located in the middle of this interval around 1~GHz. 
These Giga Hertz Peaked Sources (GPS) are powerful ($\log
P_{1.4~{\rm GHz}} > 25$~W\,Hz$^{-1}$) and compact ($< 1$~kpc)
extragalactic radio sources, which show a convex radio spectrum
peaking between 500~MHz and 10~GHz in the observer's frame (for a
review see O'Dea~1998).
The physical mechanism responsible for the turnover of the spectrum is
still unclear with two competing models proposed:
the synchrotron self-absorption caused by dense
plasma within the source or the free-free
absorption caused by a screen external to the source.
GPS sources are associated with either quasars or galaxies.
All but one objects of Fig.~\ref{fig:sibreaks} were already included in the SPECFIND V1.0 catalogue, but
the limited frequency range often prevented the detection of the spectral break. 
Four objects are included in the GPS sample of Vollmer et al. (2008) which was based on
SPECFIND V1.0 objects showing an inverted radio spectrum.

\section{Accuracy of radio catalogues \label{sec:precision}}

The cross-identification realized for SPECFIND V2.0 allowed us to evaluate the accuracy in position 
and flux density  of the 97 radio catalogues with respect to the NVSS source catalogue. 
We do not find significant systematic offset in position or flux densities between the entry 
catalogues (Table~\ref{tab:entries}) and the NVSS catalogue.

\section{Summary \label{sec:summary}}

We present the second release of the SPECFIND catalogue, SPECFIND V2.0, which is available
at CDS's VizieR\footnote{http://vizier.u-strasbg.fr/viz-bin/VizieR}. 
The catalogue contains 107488 cross-identified objects with at least 
three radio sources observed at three independent frequencies.
The cross-identification algorithm is based on proximity, source extent, survey 
resolution, flux density at the same frequency, and power law spectra at 
different frequencies (Vollmer et al. 2005a). We increased the number of
entry catalogues from 20 to 97 containing 115 tables (Table~\ref{tab:entries}). 
This large increase was only made possible
by the development of 4 tools at CDS which use the standards and infrastructure of the
Virtual Observatory (VO). This was done in the framework of the VO-TECH Design Study.
The first three tools described below are not specificly designed for radio data
and can be used for other purposes. Together with the associated manuals they
are available at http://eurovotech.org/twiki/bin/view/VOTech. These tools are:
\begin{enumerate}
\item
TABFIND: discovery of relevant resources based on Unified Content Descriptors (see Sect.~\ref{sec:tabfind}),
\item
TABUNIF: Uniformisation of a heterogeneous set of catalogues (see Sect.~\ref{sec:tabunif}),
\item
CAMEA: Description of the catalogue data (see Sect.~\ref{sec:camea}),
\item
A SPECFIND-specific tool to insure the compatibility between two successive releases
of the SPECFIND catalogue.
\end{enumerate}
With a modest increase in the number of input sources ($\sim 8$\%) we could increase the
number of cross-identified objects by $\sim 60\%$.
We decided not to remove extended and complex sources from SPECFIND V2.0 and
therefore caution the user against a blind use of the catalogue.
Typical source geometries and subsequent radio spectra are given in Sect.~\ref{sec:compatibility}).
Most frequently, the user can separate the underlying objects with the
help of the SPECFIND radio spectrum and the NVSS image.
Due to the larger frequency coverage of the SPECFIND entry catalogues with respect
to that of the previous release spectral breaks are present in the SPECFIND V2.0 catalogue.

\begin{acknowledgements}
The new tools described in the article have been developed in the frame of the VO-TECH
Design Study of the Sixth Framework Programme (Specific Support Action No 011892 
``VO-TECH  The European Virtual Observatory  VO Technology Centre'').
BV would like to thank T.~Krichbaum and W.~Reich for their precious comments
on this article.
\end{acknowledgements}


\begin{thebibliography}{}

\bibitem{a1} Altenhoff, W.J., Downes, D., Pauls, T., Schraml, J. 1979, A\&AS, 35, 23

\bibitem{a2} Altschuler, D.R. 1986, A\&AS, 65, 267

\bibitem{a3} Baldwin, J.E., Boysen, R.C., Hales, S.E.G, et al. 1985, MNRAS, 217, 717

\bibitem{a3a} Becker R.H., White R.L., \& Edwards A.L. 1991, ApJS, 75, 1

\bibitem{a4} Becker, R.H., White, R.L., Helfand, D.J., Zoonematkermani, S. 1994, ApJS, 91, 347

\bibitem{a5} Benn, C.R., Grueff, G., Vigotti, M., Wall, J.V. 1982, 200, 747 

\bibitem{a6} Benn, C.R., Kenderdine, S. 1991, MNRAS, 251, 253

\bibitem{a7} Benn, C.R. 1995, MNRAS, 272, 699 

\bibitem{a7a} Bennett A.S. 1962, MRAS, 68, 163

\bibitem{a7b} Bennett C.L., Lawrence C.R., Burke B.F., Hewitt J.N., \& Mahoney J. 1986, ApJS, 61, 1

\bibitem{a8} Bondi, M., Ciliegi, P., Zamorani, G., et al. 2003, A\&A, 403, 857

\bibitem{a8a} Bonnarel, F., Fernique, P., Bienaym\'e, O., et al. 2000, A\&AS, 143, 33

\bibitem{a8b} Browne I.W.A., Patnaik A.R., Wilkinson P.N., \& Wrobel J.M. 1998, MNRAS, 293, 257

\bibitem{a9} Ciliegi, P., McMahon, R.G., Miley, G., et al. 1999, MNRAS, 302, 222

\bibitem{a10} Cohen, A.S., Rottgering, H.J.A., Jarvis, M.J., Kassim, N.E., Lazio, T.J.W. 2004, ApJS, 150, 417

\bibitem{a11} Cohen, A.S., Lane, W.M., Cotton, W.D., et al. 2007, AJ, 134, 1245 

\bibitem{a12} Coleman, P.H., Condon, J.J., Hazard, C. 1985, AJ, 90, 1437

\bibitem{a12a} Colla G., Fanti C., Fanti R. et al. 1970, A\&AS, 1, 281 

\bibitem{a12b} Colla G., Fanti C., Fanti R. et al. 1972, A\&AS, 7, 1 

\bibitem{a12c} Colla G., Fanti C., Fanti R. et al. 1973, A\&AS, 11, 291 

\bibitem{a13} Condon, J.J., Anderson, E., Broderick, J.J. 1995, AJ, 109, 2318

\bibitem{a13a} Condon J.J., Cotton W.D., Greisen E.W. et al. 1998, AJ, 115, 1693

\bibitem{a13b} de Breuck C., Tang Y., de Bruyn A.G., Rottgering H., \& van Breugel W. 2002, A\&A, 394, 59

\bibitem{a13c} Derri\`ere S., Gray, N., McDowell, J.C., et al. 2004, in: Astronomical Data Analysis
Software and Systems (ADASS) XIII, Proceedings of the conference held 12-15 October, 2003 in 
Strasbourg, France. Edited by Francois Ochsenbein, Mark G. Allen and Daniel Egret. ASP 
Conference Proceedings, Vol. 314. San Francisco: Astronomical Society of the Pacific, 2004., p.315

\bibitem{a13d} Douglas J.N., Bash F.N., Bozyan F.A., Torrence G.W., \& Wolfe C. 1996, AJ, 111, 1945

\bibitem{a14} Dressel, L.L., Condon, J.J. 1978, ApJS, 36, 53 

\bibitem{a15} Durdin, J.M., Pleticha, D., Condon, J.J., et al. 1975, National Astronomy and Ionosphere Center Report No. 45 

\bibitem{a15a} Edge D.O., Shakeshaft J.R., McAdam W.B., Baldwin J.E., \& Archer S. 1959, MRAS, 68, 37

\bibitem{a15b} Fanti C., Fanti R., Ficarra A., \& Padrielli L. 1974, A\&AS, 18, 147

\bibitem{a16} Fanti, C., Mantovani, F., Tomasi, P. 1981, A\&AS, 43, 1

\bibitem{a17} Fanti, C., Pozzi, F., Dallacasa, D., et al. 2001, A\&A, 369, 380

\bibitem{a17a} Ficarra A., Grueff G., \& Tomassetti G. 1985, A\&AS, 59, 255

\bibitem{a18} Fich, M. 1986, AJ, 92, 787

\bibitem{a19} Filipovic, M.D., Haynes, R.F., White, G.L., et al. 1995, A\&AS, 111, 311

\bibitem{a20} Filipovic, M.D., Bohlsen, T., Reid, W., et al. 2002, MNRAS, 335, 1085

\bibitem{a21} Forkert, T.,  Altschuler, D.R. 1987, A\&AS, 70, 77

\bibitem{a21a} F\"{u}rst E., Reich W., Reich P., \& Reif K. 1990, A\&AS, 85, 805

\bibitem{a22} Garn, T., Green, D.A., Hales, S.E.G., Riley, J.M., Alexander, P. 2007, MNRAS, 376, 1251

\bibitem{a22a} Gower J.F.R., Scott P.F., \& Wills D. MRAS, 1967, 71, 49

\bibitem{a23} Green, D.A., Riley, J.M. 1995, MNRAS, 274, 324

\bibitem{a24} Gregorini, L., de Ruiter, H.R., Parma, P., et al. 1994, A\&AS, 106, 1

\bibitem{a25} Gregorini, L., Vigotti, M., Mack, K.-H., Zonnchen, J., Klein, U. 1998, A\&AS, 133, 129

\bibitem{a25a} Gregory P.C. \& Condon J.J. 1991, ApJS, 75, 1011

\bibitem{a25b} Gregory P.C., Scott W.K., Douglas K., \& Condon J.J. 1996, ApJS, 103, 427

\bibitem{a25c} Griffith M., Langston G., Heflin M. et al. 1990, ApJS, 74, 129

\bibitem{a25d} Griffith M., Langston G., Heflin M., Conner S., \& Burke B. 1991, ApJS, 75, 801

\bibitem{a25e} Griffith M.R., Wright A.E., Burke B.F., \& Ekers R.D. 1994, ApJS, 90, 179

\bibitem{a25f} Griffith M.R., Wright A.E., Burke B.F., \& Ekers R.D. 1995, ApJS, 97, 347

\bibitem{a26} Hales, S.E.G., Baldwin, J.E., Warner, P.J. 1988, MNRAS, 234, 919

\bibitem{a27} Hales, S.E.G., Masson, C.R., Warner, P.J., Baldwin, J.E. 1990, MNRAS, 246, 256

\bibitem{a28} Hales, S.E.G., Mayer, C.J., Warner, P.J., Baldwin, J.E. 1991, MNRAS, 251, 46

\bibitem{a29} Hales, S.E.G., Masson, C.R., Warner, P.J., Baldwin, J.E., Green, D.A. 1993, MNRAS, 262, 1057

\bibitem{a30} Hales, S.E.G., Waldram, E.M., Rees, N., Warner, P.J. 1995, MNRAS, 274, 447

\bibitem{a31} Hales, S.E.G., Riley, J.M., Waldram, E.M., Warner, P.J., Baldwin, J.E. 2007, MNRAS, 382, 1639

\bibitem{a32} Haynes, R.F., Caswell, J.L., Simons, L.W.J. 1979, AuJPA, 48, 1

\bibitem{a33} Hopkins, A.M., Mobasher, B., Cram, L., Rowan-Robinson, M. 1998, MNRAS, 296, 839 

\bibitem{a34} Jackson, N., Roland, J., Bremer, M., Rhee, G., Webb, J. 1999, A\&AS, 143, 401

\bibitem{a35} Jones, P.A., McAdam, W.B. 1992, 80, 137

\bibitem{a36} Kassim, N.E. 1988, ApJS, 68, 715

\bibitem{a37} Kerton, C.R., Murphy, J., Patterson, J. 2007, MNRAS, 379, 289

\bibitem{a38} Kollgaard, R.I., Brinkmann W., McMath Chester M. et al. 1994, ApJS, 93, 145

\bibitem{a39} Kulkarni, V.K., Mantovani, F., Pauliny-Toth, I.I.K. 1990, A\&AS, 2, 41

\bibitem{a39a} Landecker, T.L. \& Caswell, J.L. 1983, AJ, 88, 1810

\bibitem{a39b} Langston G.I., Heflin M.B., Conner S.R. et al. 1990, ApJS, 72, 621

\bibitem{a40} Langston, G., Minter, A., D'Addario, L., et al. 2000, AJ, 119, 2801

\bibitem{a40a} Large M.I., Cram L.E., \& Brugess A.M. 1991, The Observatory, 111, 72

\bibitem{a41} Laurent-Muehleisen, S.A., Kollgaard, R.I., Ryan, P.J., et al. 1997, A\&AS, 122, 235

\bibitem{a42} Lawrence, C.R., Bennett, C.L., Garcia-Barreto, J.A., Greenfield, P.E., Burke, B.F. 1983, ApJS, 51, 67

\bibitem{a43} Leahy, D.A., Roger, R.S. 1996, A\&AS, 115, 345

\bibitem{a44} Ledlow, M.J., Owen, F.N. 1995, AJ, 109, 853

\bibitem{a44a} Louys M., Richards A., Bonnarel, F., et al. 2008, http://www.ivoa.net/Documents/latest/CharacterisationDM.html

\bibitem{a44b} Mauch T., Murphy T., Buttery H.J. et al. 2003, MNRAS, 342, 1117

\bibitem{a45} Morganti, R. Garrett, M.A., Chapman, S., et al. 2004, A\&A, 424, 371

\bibitem{a46} Murphy, T., Mauch, T., Green, A., et al. 2007, MNRAS, 382, 382

\bibitem{a47} Myers, S.T., Jackson, N.J., Browne, I.W.A. et al. 2003, MNRAS, 341, 1 

\bibitem{a48} Niklas, S., Klein, U., Braine, J., Wielebinski, R. 1995, A\&AS, 114, 21 

\bibitem{a48a} Ochsenbein, F., Bauer, P., Marcout, J. 2000, A\&AS, 143, 23

\bibitem{q48b} O'Dea, C.P. 1998, PASP, 110, 493

\bibitem{a49} Oort, M.J.A. 1987, A\&AS, 71, 221

\bibitem{a50} Oort, M.J.A., Steemers, W.J.G., Windhorst, R.A. 1988, A\&AS, 73, 103 

\bibitem{a50a} Otrupcek R. \& Wright A.E. 1991, PASAu, 9, 1700 

\bibitem{a51} Owen, F.N., White, R.A., Hilldrup, K.C., Hanisch, R.J. 1982, AJ, 87, 1083

\bibitem{a52} Paladini, R., Burigana, C., Davies, R.D., et al. 2003, A\&A, 397, 213

\bibitem{a52a} Patnaik A.R., Browne I.W.A., Wilkinson P.N., \& Wrobel J.M. 1992, MNRAS, 254, 655

\bibitem{a53} Pearson, T.J. 1975, MNRAS, 171, 475

\bibitem{a54} Pearson, T.J. 1978, MNRAS,  182, 273

\bibitem{a55} Perley, R.A. 1982, AJ, 87, 859

\bibitem{a55a} Pilkington J.D.H. \& Scott P.F. 1965, MRAS, 69, 183

\bibitem{a56} Quiniento, Z.M., Cersosimo, J.C. 1993, A\&AS, 97, 435

\bibitem{a57} Reich, W., Furst, E., Steffen, P., Reif, K., Haslam, C.G.T. 1984, A\&AS, 58, 197

\bibitem{a58} Reich, W., Reich, P., Furst, E. 1990, A\&AS, 83, 539

\bibitem{a59} Reich, P., Reich, W., Fuerst, E. 1997, A\&AS, 126, 413

\bibitem{a60} Reich, W., Fuerst, E., Reich, P., et al. 2000, A\&A, 363, 141

\bibitem{a60a} Rengelink, R. B., Tang, Y., de Bruyn, A. G., et al. 1997, A\&AS, 124, 259

\bibitem{a61} Righetti, G., Giovannini, G., Feretti, L. 1988, A\&AS, 74, 315

\bibitem{a62} Roettgering, H.J.A., Lacy, M., Miley, G.K., Chambers, K.C., Saunders, R. 1994, A\&AS, 108, 79

\bibitem{a63} Slee, O.B. 1995, AuJPh, 48, 143 

\bibitem{a64} Slee, O.B., Perley, R.A., Siegman, B.C. 1989, AuJPh, 42, 633

\bibitem{a65} Tasse, C., Cohen, A.S., Roettgering, H.J.A., et al. 2006, A\&A, 456, 791

\bibitem{a66} Tasse, C., Roettgering, H.J.A., Best, P.N., et al. 2007, A\&A, 471, 1105

\bibitem{a67} Taylor, J.H., Manchester, R.N., Lyne, A.G. 1993, ApJS, 88, 529

\bibitem{a68} Taylor, A.R., Goss, W.M., Coleman, P.H., van Leeuwen, J., Wallace, B.J. 1996, ApJS, 107, 239

\bibitem{a69} Vollmer, B., Davoust, E., Dubois, P., et al. 2005a, A\&A, 431, 1177

\bibitem{a70} Vollmer, B., Davoust, E., Dubois, P., et al. 2005b, A\&A, 436, 757

\bibitem{a71} Vollmer, B., Krichbaum, T.P., Angelakis, E., Kovalev, Y.Y. 2008, A\&A, 489, 49

\bibitem{a72} Waldram, E.M., Pooley, G.G., Grainge, K.J.B., et al. 2003, MNRAS, 342, 915

\bibitem{a73} Walterbos, R.A.M., Brinks, E., Shane, W.W. 1985, A\&AS, 61, 451

\bibitem{a74} Wieringa, M.H. 1993, Bull. Inf. CDS 43, 17 

\bibitem{a75} Wieringa, M.H., Ekers, R.D. 2000, A\&AS, 146, 41 

\bibitem{a75a} White R.L., Becker R.H. 1992, ApJS, 79, 331

\bibitem{a75b} White R.L., Becker R.H., Helfand D.J., \& Gregg M.D. 1998, ApJ, 475, 47

\bibitem{a76} White, R.L., Becker, R.H., Helfand, D.J. 2005, AJ, 130, 586

\bibitem{a76a} Wilkinson P.N., Browne I.W.A., Patnaik A.R., Wrobel J.M., \& Sorathia B. 1998, MNRAS, 300, 790

\bibitem{a77} Windhorst, R.A., van Heerde, G.M., Katgert, P. 1984, A\&AS, 58, 1

\bibitem{a77a} Wright A.E., Griffith M.R., Burke B.F., \& Ekers R.D. 1994, ApJS, 91, 111

\bibitem{a77b} Wright A.E., Griffith M.R., Hunt A.J. et al. 1996, ApJS, 103, 145

\bibitem{a77c} Zhang X., Zheng Y., Chen H. et al. 1997, A\&AS, 121, 59

\bibitem{a78} Zhang X., Reich W., Reich P., \& Wielebinski R. 2003, A\&A, 404, 57

\bibitem{a79} Zoonematkermani, S., Helfand, D.J., Becker, R.H., White, R.L., Perley, R.A. 1990, ApJS, 74, 181 


\end{thebibliography}
\end{document}